\documentclass[aps,superscriptaddress,twocolumn]{revtex4}
\usepackage{amsfonts}
\usepackage{amsmath}
\usepackage{amssymb}
\usepackage{graphicx}

\begin{document}

\title{Optical properties of CeO$_{2}$ using screened hybrid functional and
\emph{GW}+\emph{U} methods}
\author{Hongliang Shi}
\affiliation{State Key Laboratory for Superlattices and
Microstructures, Institute of Semiconductors, Chinese Academy of
Sciences, P. O. Box 912, Beijing 100083, People's Republic of China}
\affiliation{LCP, Institute of Applied Physics and Computational
Mathematics, P.O. Box 8009, Beijing 100088, People's Republic of
China}
\author{Ping Zhang}
\thanks{Author to whom correspondence should be addressed. Electronic
address: zhang\_ping@iapcm.ac.cn} \affiliation{LCP, Institute of
Applied Physics and Computational Mathematics, P.O. Box 8009,
Beijing 100088, People's Republic of China} \affiliation{Center for
Applied Physics and Technology, Peking University, Beijing 100871,
People's Republic of China}
\author{Shu-Shen Li}
\affiliation{State Key Laboratory for Superlattices and
Microstructures, Institute of Semiconductors, Chinese Academy of
Sciences, P. O. Box 912, Beijing 100083, People's Republic of China}

\begin{abstract}
The optical spectra of CeO$_{2}$ have been systematically
investigated using three first-principles computational approaches
for comparison, namely, the Heyd-Scuseria-Ernzerhof (HSE) screened
hybrid functional, HSE+\emph{U}, and \emph{GW}+\emph{U}. Our results
show that by using the HSE+\emph{U} method, the calculated
electronic structures are in good agreement with experimental
spectra and the resulting imaginary part of the optical dielectric
function spectrum well reproduces the main features exhibited in
experimental observations. The important adsorption spectrum and
energy loss function also accord well with the experimental results.

\end{abstract}

\maketitle

Cerium oxides have been applied extensively because of the
technologically important applications in industry originating from
their unique physical properties. Taking CeO$_{2}$ for example, it
is used in automobile exhaust catalysts due to its capacity of high
oxygen storage. Another novel application of CeO$_{2}$ in the
microelectronic and optoelectronic fields has been attracted more
attention because of its high dielectric constant and good epitaxy
on Si substrates resulted from the same cubic structure and small
lattice mismatch \cite{r1}. Particularly, the dielectric function
related optical properties of CeO$_{2}$ have been investigated
intensively by several experimental work \cite{r2,r3,r4}, and the
complex 4\emph{f} orbitals of Ce are also proved to be unoccupied
locating within the band gap formed between the O $2p$ and Ce $5d$
states.

From theoretical viewpoint, accurate calculations of optical
properties for 4\emph{f} rare-earth oxide systems are hard. For
example, firstly, conventional density functional theory (DFT)
within standard local or semilocal functionals always underestimates
the band gaps of insulators or semiconductors, consequently, the
resulting optical spectra are unsatisfactory. Secondly, an accurate
description of electronic structures for rare-earth oxides
containing 4\emph{f} electrons is also a great challenge to DFT due
to the simultaneous itinerant and localized behaviors exhibited by
the \emph{f} orbitals. Many efforts have been devoted to overcoming
these intractable issues.

One successful method to remedy the drawback of the standard local
or semilocal functionals is the newly developed hybrid functionals
\cite{r5} which introduces 25\% of the nonlocal Hartree-Fock (HF)
exchange into the otherwise exact semilocal exchange functional. In
order to apply this method to large molecules or extended systems,
an improved version of the HSE screened Coulomb hybrid functional
\cite{r6} has been developed, in which the slowly decaying
long-ranged (LR) part of the HF exchange term is replaced by the
corresponding density functional counterpart. Taking
Perdew-Burke-Ernzerhof (PBE) potentials for example, the resulting
exchange is given by
\begin{equation}
E_{\mathrm{x}}^{\mathrm{HSE}}=\frac{1}{4}E_{\mathrm{x}}^{\mathrm{HF,SR}}(\mu
)+\frac{3}{4}E_{\mathrm{x}}^{\mathrm{PBE,SR}}(\mu )+E_{\mathrm{x}}^{\mathrm{%
PBE,LR}}(\mu ),
\end{equation}%
where E$_{\mathrm{x}}^{\mathrm{HF,SR}}$ is the short-ranged (SR) HF
exchange, E$_{\mathrm{x}}^{\mathrm{PBE,SR}}$ and E$_{\mathrm{x}}^{\mathrm{%
PBE,LR}}$ are the SR and LR components of the PBE exchange
functional, respectively.

Another successful method for determination of excited states is the \emph{GW%
} approximation \cite{r7}, which is based on the quasiparticle
concept
and the Green function method. In this approximation, the self-energy $%
\Sigma $ is expanded linearly in terms of the screened interaction
\emph{W}, i.e., $\Sigma\ \thickapprox\ GW$. Here, \emph{G} is the
single-particle Green's function and \emph{W} is the screened
Coulomb interaction. In order to describe the localized \emph{f}
orbitals more accurately, we also adopt the PBE plus a Hubbard
\emph{U} correction (PBE+\emph{U}) method \cite{r8} as the starting
point, which gives a qualitative improvement compared to PBE not
only for excited states such as band gap but also for ground-state
properties such as Mott insulator.

In this letter, we systematically investigate the optical spectra of
CeO$_{2} $ using the HSE, HSE+\emph{U} and \emph{GW}+\emph{U}
methods as implemented in the first-principles vienna \emph{ab
initio} simulation package (VASP) \cite{r9}. The electron and core
interactions are included using the frozen-core projected augmented
wave (PAW) approach \cite{r10}. The gradient corrected PBE
functional \cite{r11} for the exchange correlation potential is
used. The cerium 5\emph{s}, 6\emph{s}, 5\emph{p}, 5\emph{d} and
4\emph{f} as well as the oxygen 2\emph{s} and 2\emph{p} electrons
are explicitly treated as valence electrons. The electron wave
function is expanded in plane waves
up to a cutoff energy of 500 eV. For the Brillouin zone integration, the 6 $%
\times $ 6 $\times $ 6 $\Gamma $-centered \emph{k}-mesh is adopted
and a good convergence can be obtained. In the HSE calculations,
25\% HF exchange and 75 \% PBE exchange are used, and the
range-separation parameter is set to $\mu$=0.3 \AA $^{-1}$. The
\emph{GW} calculations are carried out using a total number of 166
bands and four iterations for accurate quasiparticle shifts. The
strong on-site Coulomb repulsion among the localized Ce 4\emph{f}
electrons is described by the rotationally invariant approach to the
LDA $+$ \emph{U} method due to Dudarev \emph{et al}. \cite{r12}. In
our calculation, we use \emph{J} = 0.51 eV for the exchange energy
and \emph{U} = 4.5 eV for the spherically averaged screened Coulomb
energy \emph{U}, which is close to the values used in other previous
work \cite{r13}. Since only the difference between the spherically
averaged screened Coulomb energy \emph{U} and the exchange energy
\emph{J} is important for the total LDA (GGA) energy functional
\cite{r12}, thus, in the following we label them as one single
effective parameter \emph{U} for brevity. Notice that all our
calculations are performed at the experimental lattice constants.

\begin{figure*}[ptb]
\includegraphics*[height=9.2cm,keepaspectratio]{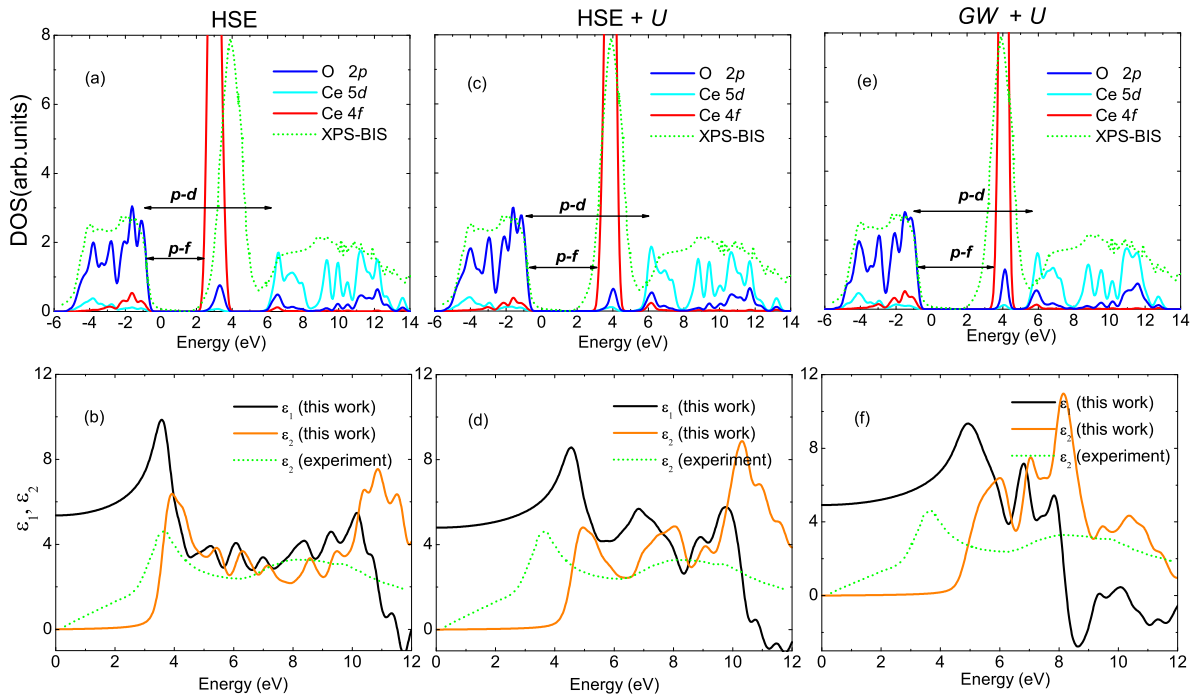}
\caption{The DOS and dielectric function spectra along with
corresponding experimental results for CeO$_{2}$: (a), (c), and (e)
the projected orbital-resolved partial DOS for O 2\emph{p}, Ce 4\emph{f} and Ce 5%
\emph{d} orbitals using HSE, HSE + \emph{U}, and \emph{GW} +
\emph{U}, respectively; (b), (d), (f) the optical spectra of the
frequency-dependent
dynamical dielectric function $\protect\varepsilon(\protect\omega)=\protect%
\varepsilon_{1}(\protect\omega)+ \emph{i}\protect\varepsilon_{2}(\protect%
\omega)$ as a function of the photon energy $\protect\omega$ using
HSE, HSE + \emph{U}, and \emph{GW} + \emph{U}, respectively. The
experimental XPS-BIS and $\varepsilon_{2}$ spectra are taken from
Refs. \cite{r16} and \cite{r2}, respectively.}
\end{figure*}

Before calculating the optical spectra of CeO$_{2}$, we first study
the essential electronic structure using the HSE, HSE+\emph{U} and \emph{%
GW}+\emph{U} methods, respectively, since the spectra are calculated
directly from the interband transitions. Based on this, the
frequency-dependent complex dielectric function $\varepsilon (\omega
)=\varepsilon _{1}(\omega )+i\varepsilon _{2}(\omega )$ for
CeO$_{2}$ are calculated. In the following, we will show and discuss
our results systematically. The calculated results obtained using
standard DFT are not present in this letter, because they can be
available elsewhere \cite{r13,r14,r15}. Figure 1(a) shows
the site-projected orbital-resolved density of states (DOS) for Ce 4\emph{%
f}, 5\emph{d} and O 2\emph{p} orbitals using HSE method together
with the experimental XPS $+$ BIS spectra \cite{r16}. From the DOS,
it is clear that the valence bands mainly consisted of O 2\emph{p}
states and a little contribution from Ce 4\emph{f} and 5\emph{d}
states can also be seen. The conduction bands mainly consisted of Ce
5\emph{d} states, and the localized Ce 4\emph{f} states are located
within the \emph{p}$-$\emph{d} gap. Our calculated width of O
2\emph{p} valence bands is about 4 eV, which is in good agreement
with the experimental XPS results \cite{r16}. Note that the
fundamental \emph{p}$-$\emph{f} and \emph{p}$-$\emph{d} gaps govern
the optical transitions, therefore, their accurate determination is
indispensable to the optical spectra. Our calculated
\emph{p}$-$\emph{f} gap is about 2.8 eV, which is well consistent
with the experimental value of 3.0 eV \cite{r16}. Furthermore, the
\emph{p}$-$\emph{d} gap of about 6.2 eV is also well determined
compared with the experimental value of about 6.0 eV \cite{r16}.
Another remarkable property is the central position of localized Ce
4\emph{f} orbitals. Our DOS showed in Fig. 1(a) shows that the
localized Ce 4\emph{f} orbitals lie about 3.5 eV above the topmost
of the O 2\emph{p} valence bands, which is smaller than the XPS-BIS
result of about 4.5 eV \cite{r16}. Notice the that present
electronic structure obtained by HSE are in good agreement with
previous study \cite{r15} using HSE.

Based on the determined electronic structure, the optical spectra of
the frequency-dependent dynamical dielectric function $\varepsilon(\omega)=%
\varepsilon_{1}(\omega)+ \emph{i}\varepsilon_{2}(\omega)$ as a
function of the photon energy $\omega$ up to 12 eV has been
calculated and showed in Fig. 1(b). In order to make a comparison,
the experimental result \cite{r2} of imaginary part
$\varepsilon_{2}(\omega)$ is also showed. Due to the cubic symmetry
of
CeO$_{2}$, the dielectric tensor only has one independent component and $%
\varepsilon_{xx}$ = $\varepsilon_{yy}$ = $\varepsilon_{zz}$, therefore, only
one component is showed in Fig. 1(b). According to our above DOS discussion,
the peak at about 3.9 eV in our calculated $\varepsilon_{2}(\omega)$ is
assigned to the transition from the O 2\emph{p} bands to the localized Ce 4%
\emph{f} states, which is close to the peak in experimental $%
\varepsilon_{2}(\omega)$ at 3.7 eV \cite{r2}. Considering the
dipolar selection rule only transitions with the difference $\Delta
l$ = $\pm$1 between the angular momentum quantum numbers \emph{l}
are allowed, the \emph{p}$-$\emph{f} transition is forbidden.
However, in our above DOS calculation, in valence bands, small
hybridization between O 2\emph{p} and Ce 4\emph{f} exists, which
suggests that some Ce 4\emph{f} electrons transfer to O 2\emph{p}
states. Notice that Niwano \emph{et al.} \cite{r3} also called the
\emph{p}$-$\emph{f}
transition as a ``charge transfer" transition. As for the O 2\emph{p}$-$Ce 5%
\emph{d} transition, experimentally Marabelli \emph{et al.}
\cite{r2} observed a
double-peaked very broad structure at about 9 eV in $\varepsilon_{2}(\omega)$%
; one peak is centered at 8 eV and the other at 10 eV. This corresponds to
the transitions from O 2\emph{p} to Ce 6\emph{d} \emph{e$_{g}$} and \emph{t$%
_{2}$} states due to crystal field splitting in fluorite crystal
structure. As for the \emph{p}$-$\emph{d} transition indicated in
$\varepsilon_{2}(\omega)$ in Fig. 1(b), one can see two peaks
located at about 8.6 and 10.8 eV, corresponding to
\emph{p}$-$\emph{d}$(e_{g})$ and \emph{p}$-$\emph{d}$(t_{2})$
transitions, respectively.

The electronic structure and optical spectra obtained by the HSE $+$ \emph{%
U} method are showed in Fig. 1(c) and Fig. 1(d). Compared to our
pure screened hybrid functional result, it is clear that the Ce
4\emph{f} states are more localized because of the Hubbard \emph{U}
correction. Furthermore, the central position of localized Ce
4\emph{f} states lies at about 4.5 eV above the topmost of the O
2\emph{p} valence bands, which becomes to accord well with the
experimental XPS-BIS measurement \cite{r16}. As for the
\emph{p}$-$\emph{f} and \emph{p}$-$\emph{d} gaps, they are now very
close to the experimental data \cite{r16}. Overall, the shape of our
calculated DOS curve exhibits the same main features as the
experimental XPS-BIS results \cite{r16}. Correspondingly, the
calculated $\varepsilon_{2}(\omega)$ by the HSE $+$ \emph{%
U} is showed in Fig. 1(d). Clearly, there are three peaks in the
calculated $\varepsilon_{2}(\omega)$. One peak is at 4.9 eV, and the
other two are at 8.1 and 10.3 eV, respectively. Comparing to the
three main peaks (located at 3.7, 8 and 10 eV) observed by
experiment \cite{r2}, we notice that the last two peaks are well
reproduced except the first peak with a shifting of about 1.2 eV
from the experimental spectrum. This discrepancy is guessed to be
due to the ignorance of the interaction between electron and hole in
our calculations. Then, the excitonic effect is not taken into
account. As Niwano \emph{et al.} \cite{r3} pointed out that the BIS
process is adding one additional electron into the empty conduction
bands, whereas optical adsorption process is exciting one electron
up to the conduction bands and a corresponding hole is created; they
further deduced the binding energy of an exciton contributed by one
hole and one 4\emph{f} electrons to be about 1 eV according to
energy difference between the central position of the localized
4\emph{f} states and the first peak in
$\varepsilon_{2}$. Notice that compared to the experimental observation \cite{r2}, our $%
\varepsilon_{2}$ spectrum also exhibits a rising edge at about 6.5
eV, which is close to the \emph{p}$-$\emph{d} gap and indicates the
beginning of the \emph{p}$-$\emph{d} transitions. Therefore, our
calculated $\varepsilon_{2}$ spectrum displays the main features as
the experimental result and supplies a qualitative agreement with
experimental spectra \cite{r2,r3}. Furthermore, if the excitonic
effect is taken into account, the oscillator strength in
$\varepsilon_{2}$ will shift to lower energies, we expect that the
agreement will become better.

The calculated projected DOS and dielectric function spectra
obtained by the \emph{GW}+\emph{U} method are also showed in Fig.
1(e) and Fig. 1(f). It is clear that the Ce 4\emph{f} orbitals are
further localized compared to the HSE+\emph{U} results. As for the
gaps, the \emph{p}$-$\emph{d} gap is close to experimental result,
whereas the \emph{p}$-$\emph{f} gap of about 3.9 eV is enlarged.
Notice that the band gaps for semiconductors or insulators are
usually overestimated in \emph{GW} scheme. This is because the
screening is reduced upon updating the eigenvalues in \emph{W}
\cite{r17}. As expected, the first peak in $\varepsilon_{2}$
spectrum will shift to a higher energy and our calculated
$\varepsilon_{2}$ spectrum shows that it is located at 6.0 eV, while
other two peaks at 8.1 and 10.3 eV, respectively.

\begin{figure}[ptb]
\includegraphics*[height=6.6cm,keepaspectratio]{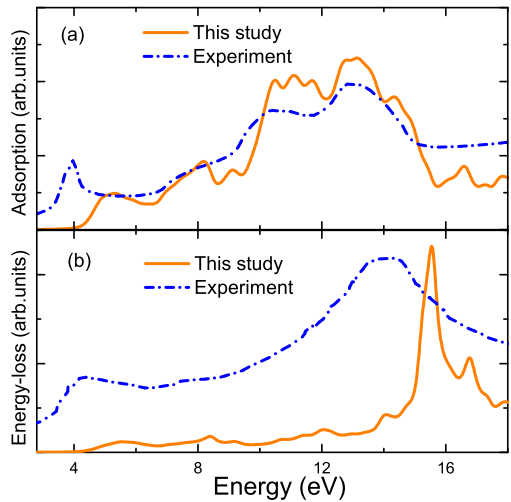}
\caption{The HSE+\emph{U} results together with corresponding
experimental results for CeO$_{2}$, (a) the optical adsorption
spectrum, (b) the energy loss function: orange solid line is this
work, blue short dash dot line is experimental results taken from
Ref. \protect\cite{r3}.}
\end{figure}

Comparing the three results showed in Fig. 1, one can conclude that
the results obtained by the HSE+\emph{U} method are better than the
other two. Furthermore, for $\varepsilon_{2}$ spectrum, we expect
the agreement will be become better if the excitonic effect is taken
into account. Based on this, the important adsorption spectrum and
energy-loss function deduced from the frequency-dependent dynamical
dielectric function obtained by HSE+\emph{U}, together with the
corresponding experimental data \cite{r3}, are showed in Fig. 2. It
is obvious that the main structures are well reproduced. Similar to
the $\varepsilon_{2}$ spectrum, the first peak in adsorption
spectrum has a shifting of about 1.2 eV to higher energy. As for the
energy-loss function, the two peaks at about 5.5 and 15.5 eV also
shift to high energies by about 1.1 eV compared to the experimental
results \cite{r3}.

In summary, we have studied the optical properties of CeO$_{2}$
using HSE, HSE+\emph{U}, and \emph{GW}+\emph{U} three methods. Using
HSE+\emph{U} method, our calculated band gaps and the central
position of 4\emph{f} orbitals are in good agreement with the
experimental values, and the resulting optical dielectric function,
adsorption spectrum and energy-loss function obtained by
HSE+\emph{U} are also well reproduced.

This work was supported by the National Basic Research Program of
China (973 Program) grant No. G2009CB929300 and the National Natural
Science Foundation of China under Grant Nos. 60521001, 60776061 and
90921003, and by the Foundations for Development of Science and
Technology of China Academy of Engineering Physics under Grant No.
2009B0301037.


\begin{references}

\bibitem{r1}
T. Inoue, Y. Yamamoto, S. Koyama, S. Suzuki, and Y. Ueda, Appl.
Phys. Lett. \textbf{56}, 1332 (1990).
\bibitem{r2}
F. Marabelli and P. Wachter, Phys. Rev. B \textbf{36}, 1238 (1987).
\bibitem{r3}
M. Niwano, S. Sato, T. Koide, T. Shidara, A. Fujimori, H. Fukutani,
S. Shin, and M. Ishigame, J. Phys. Soc. Jpn. \textbf{57}, 1489
(1988).
\bibitem{r4}
M. Veszelei, L. Kullman, C. G. Granqvist, N. Rottkay, and M. Rubin,
Appl. Opt. \textbf{37}, 5993 (1998).
\bibitem{r5}
A. D. Becke, J. Chem. Phys. \textbf{98}, 5648 (1993).
\bibitem{r6}
J. Heyd, G. Scuseria, and M. Ernzerhof, J. Chem. Phys. \textbf{118},
8207 (2003).
\bibitem{r7}
L. Hedin, Phys. Rev. \textbf{139}, A796 (1965).
\bibitem{r8}
A. Lichtenstein, J. Zaanen, and V. Anisimov, Phys. Rev. B
\textbf{52}, R5467 (1995).
\bibitem{r9}
G. Kresse and J. Hafner, Phys. Rev. B \textbf{48}, 13115 (1993).
\bibitem{r10}
P. Bl$\ddot{\rm{o}}$chl, Phys. Rev. B \textbf{50}, 17953 (1994).
\bibitem{r11}
J. P. Perdew, K. Burke, and M. Ernzerhof, Phys. Rev. Lett.
\textbf{77}, 3865 (1996).
\bibitem{r12}
S. L. Dudarev, G. A. Botton, S. Y. Savrasov, C. J. Humphreys, and A.
P. Sutton, Phys. Rev. B \textbf{57}, 1505 (1998).
\bibitem{r13}
Juarez L. F. Da Silva, M. Ver¨®nica Ganduglia-Pirovano, J. Sauer, V.
Bayer, and G. Kresse, Phys. Rev. B \textbf{75}, 045121 (2007).
\bibitem{r14}
N. Skorodumova, R. Ahuja, S. Simak, I. Abrikosov, B. Johansson, and
B. Lundqvist, Phys. Rev. B \textbf{64}, 115108 (2001).
\bibitem{r15}
P. Hay, R. Martin, J. Uddin, and G. Scuseria, J. Chem. Phys.
\textbf{125}, 034712 (2006).
\bibitem{r16}
E. Wuilloud, B. Delley, W.-D. Schneider, and Y. Baer, Phys. Rev.
Lett. \textbf{53}, 202 (1984).
\bibitem{r17}
F. Fuchs, J. Furthm¨¹ller, F. Bechstedt, M. Shishkin, and G. Kresse,
Phys. Rev. B \textbf{76}, 115109 (2007).


\end{references}
\end{document}